\documentclass[aps,superscriptaddress,prl,reprint]{revtex4-2}

\usepackage{siunitx}
\usepackage[normalem]{ulem}
\usepackage{graphicx}
\usepackage{amsmath,amssymb}
\usepackage{bm}
\usepackage[utf8]{inputenc}
\usepackage[inline]{enumitem}
\usepackage{array}
\usepackage[x11names]{xcolor}
\usepackage{comment}

\usepackage{hyperref}

\hypersetup{
  colorlinks=true,
  citecolor=blue,
  linkcolor=blue,
  urlcolor=blue
}

\usepackage{pifont}

\usepackage[capitalize]{cleveref}

\usepackage[caption=false]{subfig}

\usepackage{multirow}

\usepackage{tabularx}
\newcolumntype{Y}{>{\centering\arraybackslash}X}

\newcommand{\bmb}{BaMn$_2$Bi$_2$}

\newcommand{\ptsg}{I4^\prime/m^\prime m^\prime m}
\renewcommand{\vec}[1]{\mathbf{#1}}

\mathchardef\mhyphen="2D

\newcommand{\vecHab}{\vec{H}_{ab}}

\newcommand{\tn}{T_{\mathrm{N}}}

\newcommand{\PT}{\mathcal{PT}}
\newcommand{\Trev}{\mathcal{T}}
\newcommand{\Pinv}{\mathcal{P}}
\newcommand{\TrevAFM}{\mathcal{T}_{\mathrm{AFM}}}

\newcommand{\redsout}{\bgroup\markoverwith{\textcolor{red}{\rule[0.4ex]{4pt}{1pt}}}\ULon}

\begin{document}

\title{$\PT$-symmetry as Effective Time Reversal Symmetry for Anderson Localization \\in a Collinear Antiferromagnet}
\label{sec:authorlist}


\author{Kim-Khuong Huynh}
\email{hkkhuong@chem.au.dk}
\affiliation{Department of Chemistry, Aarhus University, Langelandsgade 140, 8000 Aarhus C, Denmark}

\author{Takuma Ogasawara}
\affiliation{BAQIS, Bld. 3, No.10 Xibeiwang East Rd., Haidian District, Beijing 100193, China}

\author{Motoi Kimata}
\affiliation{Advanced Science Research Center, Japan Atomic Energy Agency, Tokai, Ibaraki 319-1195, Japan}

\author{Sofie S{\o}by Leiszner}
\affiliation{Department of Chemistry, Aarhus University, Langelandsgade 140, 8000 Aarhus C, Denmark}

\author{Nhu-Quynh Thi Phan}
\affiliation{Department of Chemistry, Aarhus University, Langelandsgade 140, 8000 Aarhus C, Denmark}

\author{Michael Anthony Quintero}
\affiliation{Department of Chemistry, Aarhus University, Langelandsgade 140, 8000 Aarhus C, Denmark}

\author{Frej S{\o}ren Rattenborg}
\affiliation{Department of Chemistry, Aarhus University, Langelandsgade 140, 8000 Aarhus C, Denmark}

\author{Denis Ar\v{c}on}
\affiliation{Jozef Stefan Institute, Jamova cesta 39, 1000 Ljubljana, Slovenia}
\affiliation{Faculty of Mathematics and Physics, University of Ljubljana, Jadranska cesta 19, 1000 Ljubljana, Slovenia 6}

\author{Bo Brummerstedt Iversen}
\affiliation{Department of Chemistry, Aarhus University, Langelandsgade 140, 8000 Aarhus C, Denmark}

\author{Katsumi Tanigaki}
\affiliation{BAQIS, Bld. 3, No.10 Xibeiwang East Rd., Haidian District, Beijing 100193, China}


\date{\today}

\begin{abstract}
  We show that $\PT$-symmetry acts as the effective time-reversal symmetry governing Anderson localization in the antiferromagnet \bmb{}, whereas the general time-reversal symmetry is broken by its long-range magnetic order.
  The magnetoconductance follows the laws established from breaking time-reversal symmetry throuhg the orbital coupling of charge carriers to the magnet vector potential, however, the degrees of symmetry breaking are qualitatively modulated by the magnetocrystalline anisotropy stabilizing $\PT$-related sublattices.
  The quantum interferences and hoping amplitudes governing the phenomena are drastically impaired by transverse magnetic fields, which readily cant $\PT$-related magnetic sublattices.
  Accordingly, when the $\PT$-related texture is weakly perturbed by longitudinal fields orienting along the sublattices, the localization is minutely affected.
  The robustness of the $\PT$-enforced degeneracies is thus governed by exchange and magnetocrystalline interactions, in contrast to the exact $\PT$ symmetry of the full Hamiltonian, which is broken by any finite field.
\end{abstract}

\maketitle

\begin{figure}
  \centering
  \includegraphics[width = 1\linewidth]{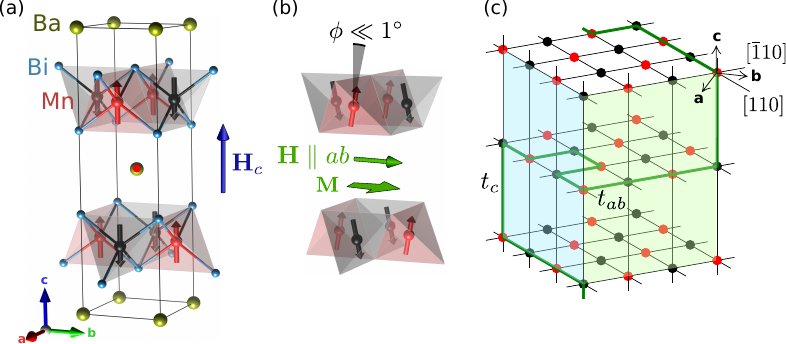}
  \caption{
    $\PT$-symmetric AFM texture in \bmb{}.
    (a) G-type AFM structure with $I4^\prime / m^\prime m^\prime m$ magnetic space group.
    Inversion $\mathcal P$ about the marked center (red dot) followed by time reversal $\mathcal T$ maps the spin-up (red) and spin-down (black) sublattices onto one another.
    A field along $\hat c$ leaves the collinear texture intact below the spin-flop field.
    (b) By contrast, an in-plane field immediately cants the moments away from $\hat c$ and removes the $\PT$ relation between the sublattices.
    (c) In the localized regime, nearest-neighbor hopping alternates between spin-up (red) and spin-down (black) Mn sites, unlike the equivalent sites of conventional paramagnetic lattice models.
  }
  \label{fig:PT_anisotropy}
\end{figure}

Spatial inversion ($\Pinv$) and time reversal ($\Trev$) are fundamental discrete symmetries governing the ordered phases and electronic states of crystalline solids.
Their spontaneous breaking underlies familiar forms of long-range order: inversion breaking permits macroscopic electric polarization in ferroelectrics, whereas time-reversal breaking accompanies magnetic order.
In a compensated collinear antiferromagnet (c-AFM), two magnetic sublattices carry equal and oppositely oriented moments, yielding zero net magnetization.
Although $\Trev$ symmetry is broken, the magnetic state may remain invariant under a combined antiunitary operation $\TrevAFM=g\Trev$, where the spatial operation $g$ interchanges the two sublattices and $\Trev$ reverses their moments.
When $g$ is spatial inversion $\Pinv$ [Fig.~\ref{fig:PT_anisotropy}(a)], $\Pinv$ exchanges the inversion-partner sublattices without reversing their axial magnetic moments, whereas $\Trev$ reverses the moments without exchanging their spatial coordinates.
Each operation separately reverses the AFM order, while their product $\TrevAFM=\Pinv\Trev$ preserves it.
Moreover, because both $\Pinv$ and $\Trev$ reverse the crystal momentum $\vec{k}$, $\Pinv\Trev$ leaves every $\vec{k}$ invariant and, for spinful electrons, satisfies $(\PT)^2=-1$ \cite{ramazashvili2008,yanase2014}.
This symmetry enforces Kramers-like double degeneracy throughout the Brillouin zone and is central to the concepts of magnetic multipole order and AFM Dirac fermions \cite{ramazashvili2008,yanase2014,watanabe2017,tang2016,smejkal2017}.



An immediate question is how this symmetry protection evolves in an applied magnetic field $\vec{H}$.
Because $\vec{H}$ is a time-odd axial vector \cite{hlinka2014}, it is invariant under $\Pinv$ but changes sign under $\Trev$; consequently, $(\PT)\,\mathcal{H}(\vec{H})\,(\PT)^{-1}=\mathcal{H}(-\vec{H})$.
Thus, the full Hamiltonian at any finite field is not $\PT$ symmetric, and its exact $\PT$-protected degeneracies are generally lifted \cite{ramazashvili2008}.
This formal symmetry breaking does not, however, determine how strongly the field reorganizes the internal AFM texture \footnote{See Supplementary Information for more details.}.
In many $\PT$-symmetric c-AFMs, exchange and spin--orbit--induced magnetocrystalline anisotropy strongly stabilize the $\PT$-related sublattice configuration against experimentally accessible Zeeman energies.
The robustness of the internal $\PT$-symmetry and the degeneracies it protects become strongly orientation dependent: the AFM texture can remain inert for one field direction yet become readily canted for another.

In this Letter, we show that $\PT$-symmetry of the magnetic texture governs the strongly localized transport in the model c-AFM \bmb{} [Fig.~\ref{fig:PT_anisotropy}(a)].
Its magnetoconductivity is uniaxial with respect to the field orientation.
It is extraordinarily sensitive to $\PT$-breaking transverse magnetic fields applied perpendicular to the easy-axis, while remaining nearly inert to longitudinal fields, which only weakly perturb the $\PT$-related AFM texture.
Measurements over multiple field orientations collapse to well-known functional depedencies on the transverse field.
These characteristic forms are well-established results of the breaking time-reversal symmetry via the coupling between the magnetic field associated vector potential and the charge degree of freedom of the conduction electrons \cite{nguyen1985,medina1989,medina1990,medina1992,medina1996,shklovskii2013}.
The emergence of these functional forms in \bmb{} demonstrates that its $\PT$-organized AFM texture plays a role analogous to that of time-reversal symmetry in conventional disordered conductors.


In a strong easy-axis c-AFM, spin--orbit--induced magnetocrystalline anisotropy pins the N\'eel vector $\vec{L}$ to a principal crystallographic axis.
In the ideal zero-temperature rigid-sublattice limit, the internal anisotropy and AFM exchange energies are much larger than the Zeeman interaction between the spins and the applied magnetic field.
Consequently, a longitudinal field, $\vec{H}_{\parallel}\parallel\vec{L}$, leaves the $\PT$-related configuration unchanged below the spin-flop field, even though the full Hamiltonian at fixed field is not $\PT$-symmetric.
By contrast, a transverse field, $\vec{H}_{\perp}\perp\vec{L}$, produces a uniform canting of the two sublattices and thereby destroys their internal $\PT$ relation.
In terms of the magnetic susceptibility, the parallel component $\chi_{\parallel}$ is much smaller than the perpendicular $\chi_{\perp}$.
As $T\rightarrow 0$, $\chi_{\parallel} \rightarrow 0$ whereas $\chi_{\perp}$ remains nearly independent of temperature \cite{yosida1996}.
The longitudinal magnetization thus vanishes, $\vec{M}_{\parallel} = \chi_{\parallel}\vec{H}_{\parallel} = 0$, and the system is inert with respect to longitudinal fields.
On the other hand, the finite transverse magnetization, $\vec{M}_{\perp} = \chi_{\perp}\vec{H}_{\perp}$, being a time-odd axial vector \cite{hlinka2014}, removes the internal $\PT$-symmetry \cite{Note1}.
If the coupling between the conduction electrons and the AFM background is strong, the electronic properties of the system is then governed by the internal $\PT$-symmetry \cite{smejkal2017,tang2016}.
The present argument applies when spin--orbit coupling cannot be neglected and locks the spin texture to the inversion-related crystallographic environments.
Remarkably, in the opposite limit of vanishing spin--orbit coupling, a transverse field can preserve a distinct hidden antiunitary symmetry and thereby retain double degeneracies at selected momenta \cite{ramazashvili2008}.
Other theoretical studies also investigated the consequences of breaking $\PT$-symmetry by longitudinal field \cite{watanabe2020,yanase2014}.

At finite temperature, $\chi_{\parallel}$ is nonzero due to the combination of Zeeman interaction and thermal excitation, but the easy-axis relation $\chi_{\parallel} \ll \chi_{\perp}$ still holds \cite{yosida1996}.
The longitudinal fields thereby may also weakly perturb the magnetic texture, nevertheless, the internal symmetry breaking remains extremely orientation dependent.
$\PT$-protected degeneracies are then dominantly lifted by the transverse field and therefore should scale with its strength, $H_{\perp} = H\sin\theta$, with $\theta$ is the angle between the total magnetic field and the easy-axis.
In the longitudinal regime, when $\vec{H}$ is nearly parallel to the easy-axis, $\PT$-breaking effects remain weak.
The angular and the magnetic field dependencies of the magnetoconductivity of \bmb{} verify this description.

\bmb{} orders magnetically below $\tn\approx\SI{387}{\kelvin}$ [Fig.~\ref{fig:PT_anisotropy}(a)] into the $\ptsg$ magnetic space group which is $\PT$-symmetric with its strong easy-axis oriented along $\hat c$ \cite{calder2014,ogasawara2021}.
The two independent component the susceptibility tensor are $\chi_{aa}$ and $\chi_{cc}$, with $\chi_{aa}\gg\chi_{cc}$ well below $\tn$ \cite{kleiner1966,seemann2015,ogasawara2021}.
All Mn $3d$ bands except $d_{xy}$ are Mott-like  with their spins ordered into the c-AFM sublattices, whereas the weakly correlated $d_{xy}$ band crosses the Fermi level \cite{craco2018,zhang2016b,jansa2021,ogasawara2022}.
Instead of a metallic ground state, \bmb{} exhibits an Anderson localization which is tunable solely by transverse magnetic field \cite{ogasawara2022}.
Although the magnetotransport properties of \bmb{} have been observed previously \cite{ogasawara2022,jansa2021,ogasawara2021,huynh2019}, their field-dependencies, which are key in estabilishing the role of $\PT$-symmetry, have not been understood until this work.

\begin{figure}
  \centering
  \includegraphics[width = .4\textwidth]{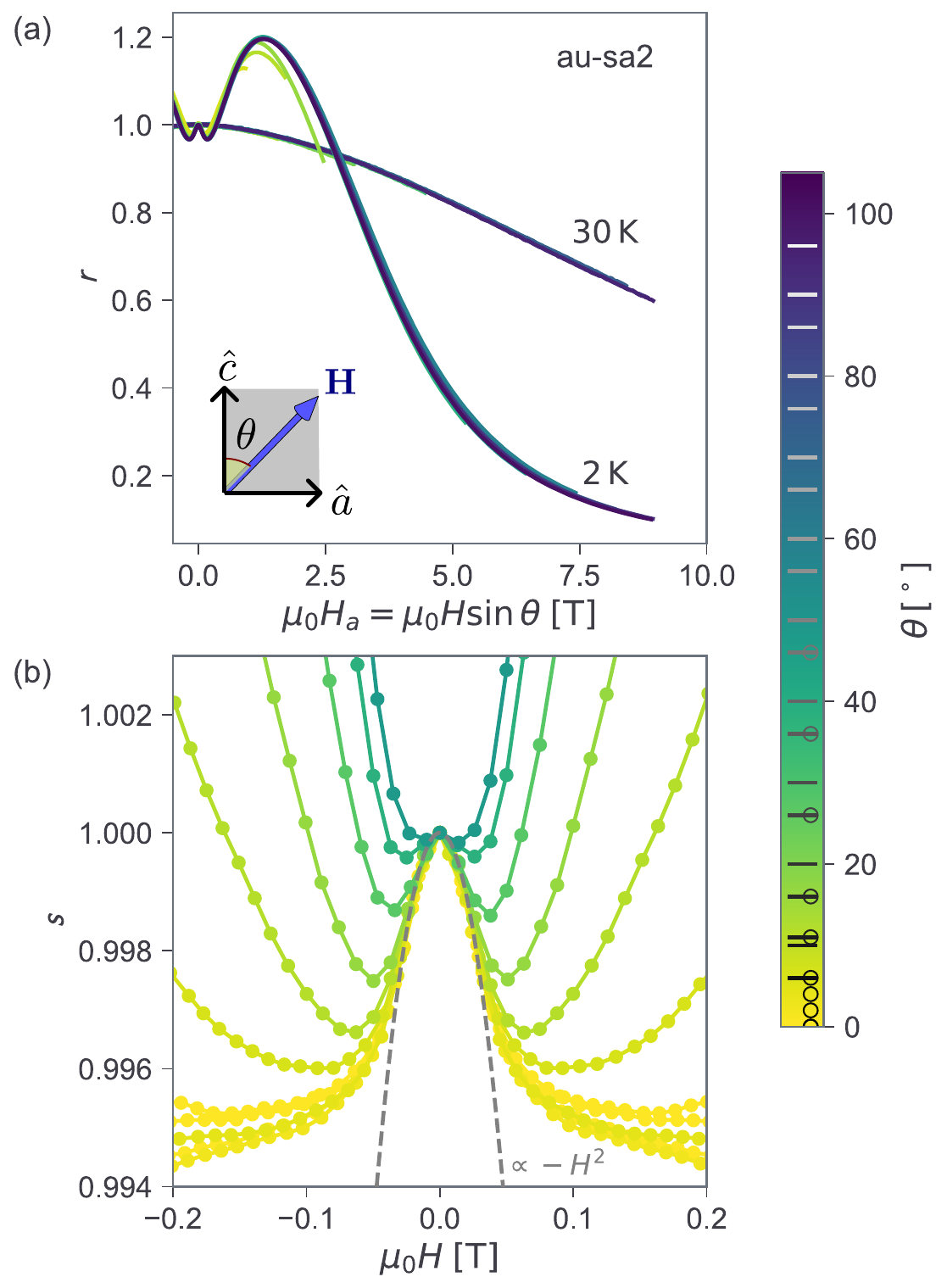}
  \caption{
    Field-induced magnetotransport in \bmb{}.
    Field orientations are defined in the inset of (a); lines and circles on the color bar identify the $\theta$-angles of the curves in (a) and (b), respectively.
    (a) The large MR induced by canting the AFM texture collapses onto a single dependence on $H_{ab}$.
    (b) At small $\theta$'s, a much smaller low-field MC cusp scales with the total field as $-H^2$.
  }
  \label{fig:MR_angles}
\end{figure}

To describe the magnetoconductivity and magnetoresistivity, we use the notations $s \equiv \sigma(\vec{H}) / \sigma(0)$ and $r \equiv \rho(\vec{H})/ \rho(0) = 1/s$, where $\sigma$ and $\rho$ are the conductivity and resistivity, respectively.
Accordingly, magnetoconductance is defined as $\mathrm{MC} = s - 1$, and magnetoresistance as $\mathrm{MR} = r -1$.
As stated earlier, the terminologies longitudinal and transverse refer to orientations of magnetic fields parallel and perpendicular to the $\hat{c}$-axis, respectively, regardless of the direction of electric current.

As shown in Fig.~\ref{fig:MR_angles}(a), $\mathrm{MR}$ of \bmb{} is extremely sensitive to the strength of the transverse field $H_{ab} = H\sin\theta$, exhibiting complex shapes with both positive and negative behaviors before saturating at about $\SI{-98}{\percent}$ in the high-$H_{ab}$ limit \cite{huynh2019,ogasawara2021,ogasawara2022}.
At each temperature, both positive and negative $\mathrm{MR}$ components of all $r$-curves at $\theta > \SI{10}{\degree}$ virtually collapse into a single dependence on $H_{ab}$.
At smaller $\theta$ angles, when the total field is nearly longitudinal, the $r$-curves gradually depart from this scaling due to a minute $\mathrm{MR}$ distinctly induced by $\vec{H}_c$.
Fig.~\ref{fig:MR_angles}(b) shows the characteristic $-{H_c}^2$ cusp displayed by $\mathrm{MC}$ in the zero-$H_c$ limit.
The $H_c$-induced $\mathrm{MR}$ never exceeds $\SI{5}{\percent}$ in the whole field range up to $H_c = \SI{18}{\tesla}$ \cite{huynh2019}, and therefore it is overwhelmed by the much larger $H_{ab}$-dependence when $\vec{H}$ orients away from the $\hat{c}$-axis.
This uniaxial dependence on the field orientation observed in the present magnetotransport phenomena holds for both intra- and inter-layer geometries of electric current, and the orientation of the electric current with respect to magnetic field is unimportant.
Rotating $\vec{H}$ in the $ab$-plane with respect to the current reveals only a minor anisotropy \cite{huynh2019,ogasawara2021}, and thus $H_a \approx H_{ab}$ in all measurements.

In understanding the observed functional dependence of the $r$-curves on $H_{ab}$, the breaking of $\Trev$-symmetry by magnetic fields in the regime of strong localization is important.
For centrosymmetric and paramagnetic conductors, the dominant $\Trev$-breaking term arises from the orbital coupling of charge carriers to the vector potential through the minimal coupling,
\begin{align}
\label{eq:A_potential}
  \mathcal{H}_{A} = \frac{1}{2m}\left(\vec{p}+e\vec{A}\right)^2 ,
\end{align}
where $-e$ is the electron charge and $\vec{A}$ is the vector potential associated with magnetic field, $\vec{B}=\boldsymbol{\nabla}\times\vec{A} = \mu_0(\vec{H} + \vec{M})$, with $\vec{M} = \bm{\chi}\vec{H}$.
Because $|\bm{\chi}|$ is small, the demagnetization correction is negligible \cite{blundell2001,Note1}.
The trajectory $\Gamma$ of a charge carrier then acquires the phase $\varphi_{\Gamma} = -\frac{e}{\hbar}\int_{\Gamma}\vec{A}d\vec{\ell}$ \cite{aharonov1959}.
The field thereby changes the relative phases and quantum interference amplitudes of electronic paths underlying the localization \cite{nguyen1985}.
In addition, the harmonic oscillator potential produced by this coupling leads to Landau quantization \cite{kittel1987}, and in the localization regime, causes the radius of the ylocalization wavefunction to shrink towards the axis of $\vec{H}$ \cite{shklovskii2013}.

Charge transport in strongly localized systems is usually modeled by multistep tunneling between the localized sites constituting a lattice [Fig.~\ref{fig:PT_anisotropy}(c)].
All possible trajectories connecting two sites occupy a characteristic ``cigar-shape'' region \cite{nguyen1985,kardar1987,medina1989,medina1990,medina1992,medina1996}.
Magnetic flux through this region changes their relative phases and partially suppresses the quantum interferences responsible for localization, yielding a large positive MC obeying the exponential scaling law $\ln s\propto H^\alpha$ with $\alpha\approx2$ \cite{nguyen1985,medina1990,medina1992,medina1996}.
Spin--orbit coupling produces saturation at high field \cite{medina1992}.
At the same time, the field-induced orbital shrinkage also reduces the hopping amplitude and results in a positive MR with the dependence $\ln r\propto H^2$ \cite{shklovskii2013}.



\begin{figure}
  \centering
  \includegraphics[width = .4\textwidth]{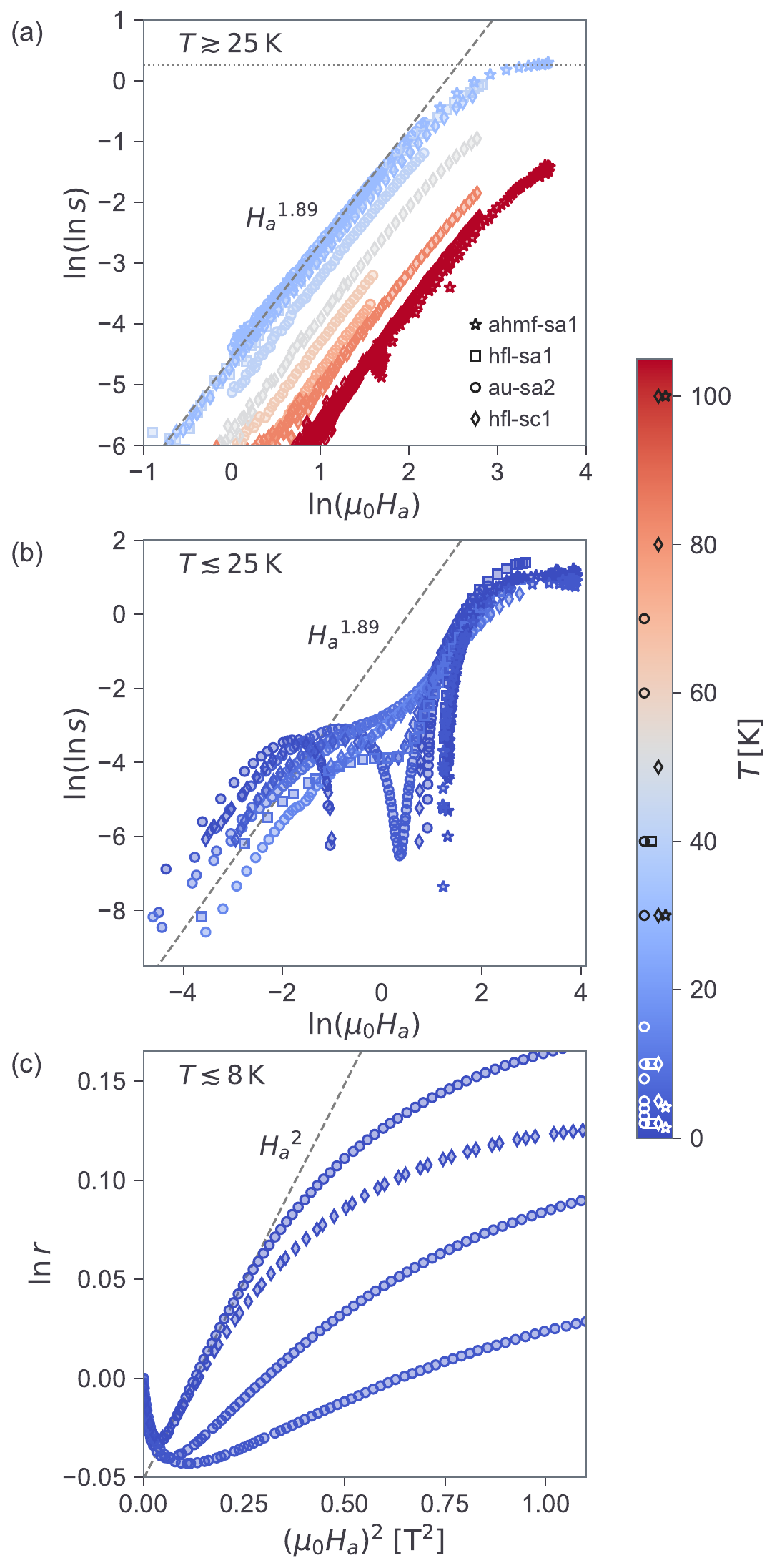}
  \caption{
    Quantum interferences and wavefunction shrinkage scaling with $H_a$.
    (a) For $T\gtrsim\SI{30}{\kelvin}$, $\ln s\propto H_a^{1.89}$ (dashed line) before saturating at high field (dotted line).
    (b) The same scaling persists for $T<\SI{30}{\kelvin}$ at low and high fields, separated by the positive-MR contribution shown in (c).
    (c) At $T<\SI{30}{\kelvin}$, the positive MR obeys $\ln r\propto H_a^2$ (dashed line).
    Marker shapes identify four samples whose details are shown in Table.~\ref{tab:sample_details}], and colors denote temperature.
    $\mu_0H_a$ is given in tesla.
  }
  \label{fig:QI}
\end{figure}

In \bmb{}, the MC retains the conventional functional forms associated with quantum interference and wavefunction shrinkage, but is governed predominantly by $H_{ab}$ rather than by the total magnetic field.
Figs.~\ref{fig:QI}(a) and \ref{fig:QI}(b) show that $\ln(\ln s)$ is linear in $\ln H_a$ with a common slope $\alpha\approx1.89$.
The scaling holds across several crystals and for currents along both $\hat a$ and $\hat c$, from the zero-field limit to high-field saturation.
Below $\SI{25}{\kelvin}$, the scaling remains visible on either side of an intervening positive-MR contribution [Fig.~\ref{fig:QI}(b)].
As shown in Fig.~\ref{fig:QI}(c), the positive MR below $\SI{25}{\kelvin}$ follows the orbital shrinkage mechanism, as evident by its $\ln r\propto H_{ab}^2$ dependence over $\SI{0.1}{\tesla}\lesssim\mu_0H_{ab}\lesssim\SI{0.5}{\tesla}$.
Finally, the minute $\mathrm{MC}\propto -{H_c}^2$ cusp observed in the longitudinal regime is consistent with an alternative localization theory based on the same coupling introduced in Eq.~\eqref{eq:A_potential} \cite{kawabata1980a}.

Thus, in \bmb{}, the localization and $\PT$-symmetry inherit the same easy-axis anisotropy: both remain weakly perturbed even in strong longitudinal fields, but respond drastically to transverse fields.
Furthermore, the $H_{ab}$-dependencies of $\mathrm{MC}$ reveal the breaking of an effective time-reversal symmetry governing both quantum interference and wavefunction shrinkage mechanisms.
Taken together, these observations identify $\PT$-symmetry as the effective time-reversal symmetry controlling Anderson localization in \bmb{}.

A complete theory of localization in a $\PT$-symmetric AFM remains to be developed, but Fig.~\ref{fig:PT_anisotropy}(c) suggests a qualitative picture.
Conventional models sum scalar amplitudes over tunnelling paths through a monochromatic lattice \cite{nguyen1985,medina1990,medina1992,medina1996,shklovskii2013}.
In a $\PT$-symmetric AFM, every nearest-neighbor path instead traverses an alternating sequence of oppositely polarized sublattices.
A complete path sum must therefore retain the spin--sublattice structure of each trajectory.
Here $\mathcal{P}$ does not neccessarilyy constrain an individual realization of the random potential.
On the other hand, together with $\mathcal{T}$, it organizes the AFM background through which the electron propagates.
Via the Zeeman interaction between the magnetic field and the AFM sublattice,
transverse-field-induced canting removes this relation between the sublattices and thereby reorganizes path interference.
In the lattice, the hoping amplitudes $t_c$ and $t_{ab}$ along $\hat{c}$-axis and in ${ab}$-plane are comparable, as supported by the small ratio between interlayer and in-plane resistivities $\rho_{c} / \rho_{ab} \approx 2$ \cite{ogasawara2021}.
We propose this as the origin of the pronounced transverse $\mathrm{MC}$; yet a quantitative formulation remains open.

The sensitivity of \bmb{} to $\PT$-breaking can be quantified by estimating the angle $\phi$ through which each $\mathrm{Mn}^{2+}$ moment cants away from the $\hat{c}$ axis under $\vecHab$.
In the collinear AFM regime, $\sin\phi = (1/2){\chi_{aa}\mu_0H_a}/{\mu_{\mathrm{Mn}^{2+}}}$ \cite{yosida1996}.
For \bmb{}, $\mu_{\mathrm{Mn}^{2+}} \approx 3.83\,\mu_{\mathrm{B}}$ and $\chi_{aa} \approx 3.8 \times 10^{-3}\,\mu_{\mathrm{B}}/(\mathrm{Mn}^{2+}\cdot\mathrm{T})$ \cite{calder2014,singh2009e,ogasawara2021}.
At $\mu_0H_a=\SI{10}{\tesla}$, where the MC is already approaching saturation, the canting angle is only $\phi\approx\SI{0.28}{\degree}$.
This corresponds to a total deviation of merely $\sim\SI{0.56}{\degree}$ from perfect antiparallel alignment between the two sublattice moments.
While the deviation from collinearity is exceedingly small, it removes $\PT$-symmetry altogether and can therefore alter the electronic properties qualitatively.
\bmb{} lies near a metal-insulator transition, and this symmetry breaking drives the system into a $\vecHab$-induced metallic state \cite{ogasawara2022}.

In summary, our experiments identify $\PT$-symmetry as the effective time-reversal symmetry governing Anderson localization in \bmb{}.
Specifically, its complex MC follows the established functional forms for quantum interference and wavefunction shrinkage when the total field $\vec{H}$ is replaced by $\vec{H}_{ab}$, the component that cants the AFM moments.
The much weaker longitudinal response arises from the same orbital coupling but without substantial reorganization of the internal $\PT$-related AFM texture; its weakness therefore reflects the strongly uniaxial magnetic response imposed by magnetocrystalline anisotropy.
This orbital route to $\PT$-symmetry breaking complements the better-known Zeeman used to probe its electronic consequences \cite{cao2023}.
The same framework may extend to other strongly anisotropic, $\PT$-symmetric c-AFM Anderson insulators, including other members of the BaMn$_2${\it Pn}$_2$ family (${\it Pn}=$ As, Sb, and Bi) and trigonal CaMn$_2$Bi$_2$, in which similarly uniaxial MC has been observed \cite{huynh2019,jansa2021,takatsu2013}.

\section*{Acknowledgements} 
The authors are grateful to M. Hagiwara, T. Kida, and T. Tahara for their excellent supports in collecting the data, to S. Kimura and S. Awaji for their hospitality and measurement support, and to Y. P. Chen for his generous instrumentation support.
K. Borup is thanked for his help in doing experiments.
Magnet-time and facilities provided by High Magnetic Field Laboratory for Superconducting Materials at Tohoku University and Center for Advanced High Magnetic Field Science at Osaka University are instrumental to this research.
The work was funded by the Villum Foundation (25861), the Danish National Research Foundation (DNRF189), and the Carlsberg Foundation (CF24-1527), by CREST project ``Thermal Management'', by Grant-in-Aid for Scientific Research on Innovative Areas ``J-Physics'' (Grant No. 18H04304), by JSPS KAKENHI (Grants numbers 18K13489, 18H03883, 17H045326, 18H03858, JP25H00599, JP25K00955, JP23KK0052, JP23H04868, and JP22H00109), by Advanced Institute for Materials Research (AIMR) at Tohoku University, by World Premier International Research Center Initiative (WPI), MEXT, Japan, and by the JSPS bilateral country research program between AIMR and Jozef Stefane Institute (Slovenia).
K.-K.H. is grateful to the Visiting Researcher's Program of the Institute for Solid State Physics, the University of Tokyo.
T.O. thanks the financial supports from the International Joint Graduate Program in Materials Science (GP-MS) of Tohoku University.
D.A. acknowledges the financial support of the Slovenian Research Agency through P1-0125 grant.
T.O. and K.T. acknowledge the Innovation Program for Quantum Science and Technology (Grant No. 2023ZD0300500) and Beijing Natural Science Foundation (Grant No. IS25007).

\section*{Author contribution}
K.-K.H. conceptualized the project, analyzed and intepreted the magnetotransport properties and wrote the orginal manuscript.
T.O. and K.-K.H. synthesized the single crystals.
T.O., M.K., and K.-K.H. investigated the magnetotransport properties and collect the data.
T.O. and N.-Q.T.P. confirmed the chemical composition of the samples by energy dispersive X-ray spectroscopy.
S.S.L, M.A.Q, and F.S.R. investigated the crystal structure by single crystal X-ray diffraction.
D.A, B.B.I, and K.T. peformed formal analyses, contributed to interpretation, and edited the paper.
B.B.I. and K.T. secured the fundings.
All authors contributed to writing, editing and reviewing the paper.

\bibliographystyle{apsrev4-2}
\bibliography{./BMPn_bibtex.bib}

@article{aharonov1959,
  title = {Significance of {{Electromagnetic Potentials}} in the {{Quantum Theory}}},
  author = {Aharonov, Y. and Bohm, D.},
  year = 1959,
  month = aug,
  journal = {Physical Review},
  volume = {115},
  number = {3},
  pages = {485--491},
  doi = {10.1103/PhysRev.115.485},
  url = {http://link.aps.org/doi/10.1103/PhysRev.115.485},
  urldate = {2014-02-20}
}

@book{blundell2001,
  title = {Magnetism in Condensed Matter},
  author = {Blundell, Stephen},
  year = 2001,
  publisher = {Oxford Univ. Press}
}

@article{calder2014,
  title = {Magnetic Structure and Spin Excitations in \$\textbraceleft\textbackslash text\textbraceleft{{BaMn}}\textbraceright\textbraceright\_\textbraceleft 2\textbraceright\textbraceleft\textbackslash text\textbraceleft{{Bi}}\textbraceright\textbraceright\_\textbraceleft 2\textbraceright\$},
  author = {Calder, S. and Saparov, B. and Cao, H. B. and Niedziela, J. L. and Lumsden, M. D. and Sefat, A. S. and Christianson, A. D.},
  year = 2014,
  month = feb,
  journal = {Physical Review B},
  volume = {89},
  number = {6},
  pages = {064417},
  publisher = {American Physical Society},
  doi = {10.1103/PhysRevB.89.064417},
  url = {https://link.aps.org/doi/10.1103/PhysRevB.89.064417},
  urldate = {2021-04-08}
}

@article{cano2017,
  title = {Chiral Anomaly Factory: {{Creating Weyl}} Fermions with a Magnetic Field},
  shorttitle = {Chiral Anomaly Factory},
  author = {Cano, Jennifer and Bradlyn, Barry and Wang, Zhijun and Hirschberger, Max and Ong, N. P. and Bernevig, B. A.},
  year = 2017,
  month = apr,
  journal = {Physical Review B},
  volume = {95},
  number = {16},
  pages = {161306},
  publisher = {American Physical Society},
  doi = {10.1103/PhysRevB.95.161306},
  url = {https://link.aps.org/doi/10.1103/PhysRevB.95.161306},
  urldate = {2021-11-12}
}

@article{cao2023,
  title = {In-{{Plane Anomalous Hall Effect}} in \$\textbackslash mathcal\textbraceleft{{PT}}\textbraceright\$-{{Symmetric Antiferromagnetic Materials}}},
  author = {Cao, Jin and Jiang, Wei and Li, Xiao-Ping and Tu, Daifeng and Zhou, Jiadong and Zhou, Jianhui and Yao, Yugui},
  year = 2023,
  month = apr,
  journal = {Physical Review Letters},
  volume = {130},
  number = {16},
  pages = {166702},
  publisher = {American Physical Society},
  doi = {10.1103/PhysRevLett.130.166702},
  url = {https://link.aps.org/doi/10.1103/PhysRevLett.130.166702},
  urldate = {2023-09-01}
}

@article{craco2018,
  title = {Microscopic Description of Orbital-Selective Spin Ordering in \$\textbraceleft\textbackslash mathrm\textbraceleft{{BaMn}}\textbraceright\textbraceright\_\textbraceleft 2\textbraceright\textbraceleft\textbackslash mathrm\textbraceleft{{As}}\textbraceright\textbraceright\_\textbraceleft 2\textbraceright\$},
  author = {Craco, L. and Carara, S. S.},
  year = 2018,
  month = may,
  journal = {Physical Review B},
  volume = {97},
  number = {20},
  pages = {205114},
  publisher = {American Physical Society},
  doi = {10.1103/PhysRevB.97.205114},
  url = {https://link.aps.org/doi/10.1103/PhysRevB.97.205114},
  urldate = {2021-11-11}
}

@misc{Crysalis,
  title = {{{CrysAlis}}{{{\textsuperscript{Pro}}}}},
  author = {Diffraction, Rigaku Oxford},
  year = 2025,
  address = {Yarnton, Oxfordshire, England},
  howpublished = {Rigaku Oxford Diffraction/Agilent Technologies Ltd.}
}

@article{dolomanov2009,
  title = {{{OLEX2}}: A Complete Structure Solution, Refinement and Analysis Program},
  shorttitle = {{{OLEX2}}},
  author = {Dolomanov, O. V. and Bourhis, L. J. and Gildea, R. J. and Howard, J. a. K. and Puschmann, H.},
  year = 2009,
  month = apr,
  journal = {Journal of Applied Crystallography},
  volume = {42},
  number = {2},
  pages = {339--341},
  publisher = {International Union of Crystallography},
  issn = {0021-8898},
  doi = {10.1107/S0021889808042726},
  url = {https://journals.iucr.org/j/issues/2009/02/00/kk5042/},
  urldate = {2026-03-20},
  langid = {english}
}

@article{hikami1980,
  title = {Spin-{{Orbit Interaction}} and {{Magnetoresistance}} in the {{Two Dimensional Random System}}},
  author = {Hikami, Shinobu and Larkin, Anatoly I. and Nagaoka, Yosuke},
  year = 1980,
  month = jan,
  journal = {Progress of Theoretical Physics},
  volume = {63},
  number = {2},
  pages = {707--710},
  issn = {0033-068X, 1347-4081},
  doi = {10.1143/PTP.63.707},
  url = {http://ptp.oxfordjournals.org/content/63/2/707},
  urldate = {2013-07-20},
  langid = {english}
}

@article{hlinka2014,
  title = {Eight {{Types}} of {{Symmetrically Distinct Vectorlike Physical Quantities}}},
  author = {Hlinka, J.},
  year = 2014,
  month = oct,
  journal = {Physical Review Letters},
  volume = {113},
  number = {16},
  pages = {165502},
  publisher = {American Physical Society},
  doi = {10.1103/PhysRevLett.113.165502},
  url = {https://link.aps.org/doi/10.1103/PhysRevLett.113.165502},
  urldate = {2023-08-29}
}

@article{huynh2019,
  title = {Negative and positive magnetoresistance in the itinerant antiferromagnet BaMn{$_2$}Pn{$_2$} ( Pn = P, As, Sb, and Bi)},
  author = {Huynh, Kim-Khuong and Ogasawara, Takuma and Kitahara, Keita and Tanabe, Yoichi and Matsushita, Stephane Yu and Tahara, Time and Kida, Takanori and Hagiwara, Masayuki and Arčon, Denis and Tanigaki, Katsumi},
  year = 2019,
  month = may,
  journal = {Physical Review B},
  volume = {99},
  number = {19},
  pages = {195111},
  issn = {2469-9950, 2469-9969},
  doi = {10.1103/PhysRevB.99.195111},
  url = {https://link.aps.org/doi/10.1103/PhysRevB.99.195111},
  urldate = {2019-05-07},
  langid = {english}
}

@article{jansa2021,
  title = {Electron correlations and charge segregation in layered manganese pnictide antiferromagnets showing anomalously large magnetoresistance},
  author = {Janša, N. and Huynh, K.-K. and Ogasawara, T. and Klanjšek, M. and Jeglič, P. and Carretta, P. and Tanigaki, K. and Arčon, D.},
  year = 2021,
  month = feb,
  journal = {Physical Review B},
  volume = {103},
  number = {6},
  pages = {064422},
  publisher = {American Physical Society},
  doi = {10.1103/PhysRevB.103.064422},
  url = {https://link.aps.org/doi/10.1103/PhysRevB.103.064422},
  urldate = {2021-02-24}
}

@article{kardar1987,
  title = {Replica {{Bethe}} Ansatz Studies of Two-Dimensional Interfaces with Quenched Random Impurities},
  author = {Kardar, Mehran},
  year = 1987,
  month = jan,
  journal = {Nuclear Physics B},
  volume = {290},
  pages = {582--602},
  issn = {0550-3213},
  doi = {10.1016/0550-3213(87)90203-3},
  url = {https://www.sciencedirect.com/science/article/pii/0550321387902033},
  urldate = {2024-10-09}
}

@article{kawabata1980a,
  title = {Theory of Negative Magnetoresistance in Three-Dimensional Systems},
  author = {Kawabata, A.},
  year = 1980,
  month = may,
  journal = {Solid State Communications},
  volume = {34},
  number = {6},
  pages = {431--432},
  issn = {0038-1098},
  doi = {10.1016/0038-1098(80)90644-4},
  url = {http://www.sciencedirect.com/science/article/pii/0038109880906444},
  urldate = {2013-10-24}
}

@book{kittel1987,
  title = {Quantum {{Theory}} of {{Solids}}},
  author = {Kittel, Charles and {Ching-yao Fong}},
  year = 1987,
  month = apr,
  publisher = {Wiley},
  googlebooks = {BcGwQgAACAAJ},
  isbn = {978-0-471-62412-7},
  langid = {english}
}

@article{kleiner1966,
  title = {Space-{{Time Symmetry}} of {{Transport Coefficients}}},
  author = {Kleiner, W. H.},
  year = 1966,
  month = feb,
  journal = {Physical Review},
  volume = {142},
  number = {2},
  pages = {318--326},
  publisher = {American Physical Society},
  doi = {10.1103/PhysRev.142.318},
  url = {https://link.aps.org/doi/10.1103/PhysRev.142.318},
  urldate = {2023-09-01}
}

@article{medina1989,
  title = {Interference of {{Directed Paths}} in {{Disordered Systems}}},
  author = {Medina, Ernesto and Kardar, Mehran and Shapir, Yonathan and Wang, Xiang Rong},
  year = 1989,
  month = feb,
  journal = {Physical Review Letters},
  volume = {62},
  number = {8},
  pages = {941--944},
  publisher = {American Physical Society},
  doi = {10.1103/PhysRevLett.62.941},
  url = {https://link.aps.org/doi/10.1103/PhysRevLett.62.941},
  urldate = {2020-09-24}
}

@article{medina1990,
  title = {Magnetic-Field Effects on Strongly Localized Electrons},
  author = {Medina, Ernesto and Kardar, Mehran and Shapir, Yonathan and Wang, Xiang Rong},
  year = 1990,
  month = apr,
  journal = {Physical Review Letters},
  volume = {64},
  number = {15},
  pages = {1816--1819},
  publisher = {American Physical Society},
  doi = {10.1103/PhysRevLett.64.1816},
  url = {https://link.aps.org/doi/10.1103/PhysRevLett.64.1816},
  urldate = {2021-01-05}
}

@article{medina1992,
  title = {Quantum Interference Effects for Strongly Localized Electrons},
  author = {Medina, Ernesto and Kardar, Mehran},
  year = 1992,
  month = oct,
  journal = {Physical Review B},
  volume = {46},
  number = {16},
  pages = {9984--10006},
  publisher = {American Physical Society},
  doi = {10.1103/PhysRevB.46.9984},
  url = {https://link.aps.org/doi/10.1103/PhysRevB.46.9984},
  urldate = {2024-08-26}
}

@article{medina1996,
  title = {Magnetoconductance Anisotropy and Interference Effects in Variable-Range Hopping},
  author = {Medina, Ernesto and Kardar, Mehran and Rangel, Rafael},
  year = 1996,
  month = mar,
  journal = {Physical Review B},
  volume = {53},
  number = {12},
  pages = {7663--7672},
  publisher = {American Physical Society},
  doi = {10.1103/PhysRevB.53.7663},
  url = {https://link.aps.org/doi/10.1103/PhysRevB.53.7663},
  urldate = {2020-09-24}
}

@article{nguyen1985,
  title = {Tunnel Hopping in Disordered Systems},
  author = {Nguyen, V L and Spivak, B Z and Shklovskii, B I},
  year = 1985,
  month = nov,
  journal = {Journal of Experimental and Theoretical Physics},
  volume = {62},
  number = {5},
  pages = {1021--1029},
  langid = {english}
}

@article{ogasawara2021,
  title = {Large negative magnetoresistance in the antiferromagnet {${\mathrm{BaMn}}_{2}{\mathrm{Bi}}_{2}$}},
  author = {Ogasawara, Takuma and Huynh, Kim-Khuong and Tahara, Time and Kida, Takanori and Hagiwara, Masayuki and Arčon, Denis and Kimata, Motoi and Matsushita, Stephane Yu and Nagata, Kazumasa and Tanigaki, Katsumi},
  year = 2021,
  month = mar,
  journal = {Physical Review B},
  volume = {103},
  number = {12},
  pages = {125108},
  publisher = {American Physical Society},
  doi = {10.1103/PhysRevB.103.125108},
  url = {https://link.aps.org/doi/10.1103/PhysRevB.103.125108},
  urldate = {2021-03-05}
}

@article{ogasawara2022,
  title = {Magnetic field induced {Anderson} localization in the orbital-selective antiferromagnet ${\mathrm{BaMn}}_{2}{\mathrm{Bi}}_{2}$},
  author = {Ogasawara, Takuma and Huynh, Kim-Khuong and Matsushita, Stephane Yu and Kimata, Motoi and Tahara, Time and Kida, Takanori and Hagiwara, Masayuki and Arčon, Denis and Tanigaki, Katsumi},
  year = 2022,
  month = jul,
  journal = {Physical Review B},
  volume = {106},
  number = {4},
  pages = {L041114},
  publisher = {American Physical Society},
  doi = {10.1103/PhysRevB.106.L041114},
  url = {https://link.aps.org/doi/10.1103/PhysRevB.106.L041114},
  urldate = {2022-07-28}
}

@article{ramazashvili2008,
  title = {Kramers {{Degeneracy}} in a {{Magnetic Field}} and {{Zeeman Spin-Orbit Coupling}} in {{Antiferromagnetic Conductors}}},
  author = {Ramazashvili, Revaz},
  year = 2008,
  month = sep,
  journal = {Physical Review Letters},
  volume = {101},
  number = {13},
  pages = {137202},
  publisher = {American Physical Society},
  doi = {10.1103/PhysRevLett.101.137202},
  url = {https://link.aps.org/doi/10.1103/PhysRevLett.101.137202},
  urldate = {2026-08-19}
}

@article{saparov2013,
  title = {Crystals, Magnetic and Electronic Properties of a New {{ThCr2Si2-type BaMn2Bi2}} and {{K-doped}} Compositions},
  author = {Saparov, Bayrammurad and Sefat, Athena S.},
  year = 2013,
  month = aug,
  journal = {Journal of Solid State Chemistry},
  volume = {204},
  pages = {32--39},
  issn = {0022-4596},
  doi = {10.1016/j.jssc.2013.05.010},
  url = {https://www.sciencedirect.com/science/article/pii/S0022459613002545},
  urldate = {2021-09-02},
  langid = {english}
}

@article{seemann2015,
  title = {Symmetry-Imposed Shape of Linear Response Tensors},
  author = {Seemann, M. and K{\"o}dderitzsch, D. and Wimmer, S. and Ebert, H.},
  year = 2015,
  month = oct,
  journal = {Physical Review B},
  volume = {92},
  number = {15},
  pages = {155138},
  publisher = {American Physical Society},
  doi = {10.1103/PhysRevB.92.155138},
  url = {https://link.aps.org/doi/10.1103/PhysRevB.92.155138},
  urldate = {2023-09-01}
}

@article{sheldrick2015,
  title = {{{SHELXT}} -- {{Integrated}} Space-Group and Crystal-Structure Determination},
  author = {Sheldrick, G. M.},
  year = 2015,
  month = jan,
  journal = {Acta Crystallographica Section A: Foundations and Advances},
  volume = {71},
  number = {1},
  pages = {3--8},
  publisher = {International Union of Crystallography},
  issn = {2053-2733},
  doi = {10.1107/S2053273314026370},
  url = {//journals.iucr.org/paper?sc5086},
  urldate = {2026-03-20},
  langid = {english}
}

@article{sheldrick2015a,
  title = {Crystal Structure Refinement with {{SHELXL}}},
  author = {Sheldrick, G. M.},
  year = 2015,
  month = jan,
  journal = {Acta Crystallographica Section C: Structural Chemistry},
  volume = {71},
  number = {1},
  pages = {3--8},
  publisher = {International Union of Crystallography},
  issn = {2053-2296},
  doi = {10.1107/S2053229614024218},
  url = {//journals.iucr.org/paper?fa3356},
  urldate = {2026-03-20},
  langid = {english}
}

@book{shklovskii2013,
  title = {Electronic properties of doped semiconductors},
  author = {Shklovskii, Boris Isaakovich and Efros, Alex L},
  year = 2013,
  volume = {45},
  publisher = {Springer Science \& Business Media}
}

@article{singh2009e,
  title = {Magnetic order in BaMn_{2}As_{2} from neutron diffraction measurements},
  author = {Singh, Yogesh and Green, M. A. and Huang, Q. and Kreyssig, A. and McQueeney, R. J. and Johnston, D. C. and Goldman, A. I.},
  year = 2009,
  month = sep,
  journal = {Physical Review B},
  volume = {80},
  number = {10},
  pages = {100403},
  doi = {10.1103/PhysRevB.80.100403},
  url = {http://link.aps.org/doi/10.1103/PhysRevB.80.100403},
  urldate = {2013-04-04}
}

@article{smejkal2017,
  title = {Route towards {{Dirac}} and {{Weyl}} Antiferromagnetic Spintronics},
  author = {{\v S}mejkal, Libor and Jungwirth, Tom{\'a}{\v s} and Sinova, Jairo},
  year = 2017,
  journal = {physica status solidi (RRL) -- Rapid Research Letters},
  volume = {11},
  number = {4},
  pages = {1700044},
  issn = {1862-6270},
  doi = {10.1002/pssr.201700044},
  url = {https://onlinelibrary.wiley.com/doi/abs/10.1002/pssr.201700044},
  urldate = {2021-11-14},
  langid = {english}
}

@article{takatsu2013,
  title = {Extremely {{Large Magnetoresistance}} in the {{Nonmagnetic Metal PdCoO}} 2},
  author = {Takatsu, Hiroshi and Ishikawa, Jun J. and Yonezawa, Shingo and Yoshino, Harukazu and Shishidou, Tatsuya and Oguchi, Tamio and Murata, Keizo and Maeno, Yoshiteru},
  year = 2013,
  month = aug,
  journal = {Physical Review Letters},
  volume = {111},
  number = {5},
  issn = {0031-9007, 1079-7114},
  doi = {10.1103/PhysRevLett.111.056601},
  url = {https://link.aps.org/doi/10.1103/PhysRevLett.111.056601},
  urldate = {2018-09-12},
  langid = {english}
}

@article{tang2016,
  title = {Dirac Fermions in an Antiferromagnetic Semimetal},
  author = {Tang, Peizhe and Zhou, Quan and Xu, Gang and Zhang, Shou-Cheng},
  year = 2016,
  month = dec,
  journal = {Nature Physics},
  volume = {12},
  number = {12},
  pages = {1100--1104},
  publisher = {Nature Publishing Group},
  issn = {1745-2481},
  doi = {10.1038/nphys3839},
  url = {https://www.nature.com/articles/nphys3839},
  urldate = {2020-12-24},
  copyright = {2016 Nature Publishing Group},
  langid = {english}
}

@article{watanabe2017,
  title = {Magnetic Hexadecapole Order and Magnetopiezoelectric Metal State in \$\textbraceleft\textbackslash mathbf\textbraceleft{{Ba}}\textbraceright\textbraceright\_\textbraceleft 1\textbackslash ensuremath\textbraceleft -\textbraceright x\textbraceright\textbraceleft\textbackslash mathbf\textbraceleft{{K}}\textbraceright\textbraceright\_\textbraceleft x\textbraceright\textbraceleft\textbackslash mathbf\textbraceleft{{Mn}}\textbraceright\textbraceright\_\textbraceleft 2\textbraceright\textbraceleft\textbackslash mathbf\textbraceleft{{As}}\textbraceright\textbraceright\_\textbraceleft 2\textbraceright\$},
  author = {Watanabe, Hikaru and Yanase, Youichi},
  year = 2017,
  month = aug,
  journal = {Physical Review B},
  volume = {96},
  number = {6},
  pages = {064432},
  publisher = {American Physical Society},
  doi = {10.1103/PhysRevB.96.064432},
  url = {https://link.aps.org/doi/10.1103/PhysRevB.96.064432},
  urldate = {2021-10-22}
}

@article{watanabe2020,
  title = {Nonlinear Electric Transport in Odd-Parity Magnetic Multipole Systems: {{Application}} to {{Mn-based}} Compounds},
  shorttitle = {Nonlinear Electric Transport in Odd-Parity Magnetic Multipole Systems},
  author = {Watanabe, Hikaru and Yanase, Youichi},
  year = 2020,
  month = oct,
  journal = {Physical Review Research},
  volume = {2},
  number = {4},
  pages = {043081},
  publisher = {American Physical Society},
  doi = {10.1103/PhysRevResearch.2.043081},
  url = {https://link.aps.org/doi/10.1103/PhysRevResearch.2.043081},
  urldate = {2021-08-13}
}

@article{yanase2014,
  title = {Magneto-{{Electric Effect}} in {{Three-Dimensional Coupled Zigzag Chains}}},
  author = {Yanase, Youichi},
  year = 2014,
  month = jan,
  journal = {Journal of the Physical Society of Japan},
  volume = {83},
  number = {1},
  pages = {014703},
  publisher = {The Physical Society of Japan},
  issn = {0031-9015},
  doi = {10.7566/JPSJ.83.014703},
  url = {https://journals.jps.jp/doi/10.7566/JPSJ.83.014703},
  urldate = {2022-08-01}
}

@book{yosida1996,
  title = {Theory of {{Magnetism}}},
  author = {Yosida, Kei},
  year = 1996,
  volume = {122},
  publisher = {Springer Science \& Business Media},
  isbn = {3-540-60651-3}
}

@article{zhang2016b,
  title = {Angle-Resolved Photoemission Observation of {{Mn-pnictide}} Hybridization and Negligible Band Structure Renormalization in \$\textbraceleft\textbackslash mathrm\textbraceleft{{BaMn}}\textbraceright\textbraceright\_\textbraceleft 2\textbraceright\textbraceleft\textbackslash mathrm\textbraceleft{{As}}\textbraceright\textbraceright\_\textbraceleft 2\textbraceright\$ and \$\textbraceleft\textbackslash mathrm\textbraceleft{{BaMn}}\textbraceright\textbraceright\_\textbraceleft 2\textbraceright\textbraceleft\textbackslash mathrm\textbraceleft{{Sb}}\textbraceright\textbraceright\_\textbraceleft 2\textbraceright\$},
  author = {Zhang, W.-L. and Richard, P. and {\noopsort{roekeghem}}{van Roekeghem}, A. and Nie, S.-M. and Xu, N. and Zhang, P. and Miao, H. and Wu, S.-F. and Yin, J.-X. and Fu, B. B. and Kong, L.-Y. and Qian, T. and Wang, Z.-J. and Fang, Z. and Sefat, A. S. and Biermann, S. and Ding, H.},
  year = 2016,
  month = oct,
  journal = {Physical Review B},
  volume = {94},
  number = {15},
  pages = {155155},
  publisher = {American Physical Society},
  doi = {10.1103/PhysRevB.94.155155},
  url = {https://link.aps.org/doi/10.1103/PhysRevB.94.155155},
  urldate = {2021-09-01}
}

@preamble{ "\providecommand{\noopsort}[1]{} " }

\clearpage
\appendix

\label{sec:endmater}

The details of synthesis, chemical composition, and structure of \bmb{} crystals are described in the Supplementary Information \cite{Note1}.
The magnetotransport properties of multiple \bmb{} crystals were measured at AIMR, High Field Laboratory for Superconducting Materials at Tohoku University (HFLSM), Center for Advanced High Magnetic Field Science at Osaka University (AHMF), and Departmetn of Chemistry, Aarhus University (AU).
The shapes of the $r$-curves observed from all measurements are consistent to that reported in Fig.~\ref{fig:MR_angles} and Fig. \ref{fig:QI}.
Parts of the collected data were previously published \cite{ogasawara2022,jansa2021,ogasawara2021,huynh2019}, but the breaking of $\PT$-symmetry had not been realized until this work.
The sample naming convention and the measurement details of the data displayed in Fig.~\ref{fig:MR_angles} and Fig. \ref{fig:QI} are shown in Table \ref{tab:sample_details}.

In the preparation of transport samples, single crystals of \bmb{} were cleaved to expose fresh $ab$ surfaces.
The crystals were then cut into rectangular shapes whose edges were aligned along the $\hat{a}$- and $\hat{b}$-axes.
Six electrodes were then painted on the sample by DuPont Ag-paste to allow an electric current to flow either along $\hat{a}$- or $\hat{c}$-axis while magnetoresistance and Hall effects could be measured.
The dimensions of the samples used for in-plane measurements were typically of about $\SI{1}\times\SI{.5}\times\SI{.05}{\cubic \milli \meter}$, whereas those of the $\hat{c}$-axis transport samples were about $\SI{.5}\times\SI{.5}\times\SI{1.5}{\cubic \milli \meter}$.
Due to the sensitivity to air of the compound, care was taken so that the samples were exposed less than $\SI{2}{\hour}$ in the air.

The effects of electrode misalignments in the magnetoresistance curve were eliminated by averaging the $H$ and $-H$ data.
The angular dependencies of $\mathrm{MR}$ were collected with the help of a Quantum Design horizontal rotator and home-built dual-axis rotators.

\begin{table}[!ht]
  \caption{Measurement details.}
  \label{tab:sample_details}
  \small
  \begin{tabular}{ccl}
    \hline
    ahmf-sa1 &
    $\vec{j}\parallel\hat{a}$ &
    \shortstack[l]{\SI{50}{\tesla} pulsed field, AHMF} \\

    hlf-sa1 &
    $\vec{j}\parallel\hat{a}$ &
    \shortstack[l]{\SI{18}{\tesla} superconducting magnet, HFLSM} \\

    hfl-sc1 &
    $\vec{j}\parallel\hat{c}$ &
    As above \\

    au-sa2 &
    $\vec{j}\parallel\hat{a}$ &
    \shortstack[l]{\SI{9}{\tesla} Quantum Design PPMS, AU} \\
    \hline
  \end{tabular}
\end{table}

\end{document}


\title{$\mathcal{PT}$-symmetry as Effective Time Reversal for Anderson Localization \\in a Collinear Antiferromagnet\\-- Suplementary Information}
\label{sec:authorlist}


\author{Kim-Khuong Huynh}
\email{hkkhuong@chem.au.dk}
\affiliation{Department of Chemistry, Aarhus University, Langelandsgade 140, 8000 Aarhus C, Denmark}

\author{Takuma Ogasawara}
\affiliation{BAQIS, Bld. 3, No.10 Xibeiwang East Rd., Haidian District, Beijing 100193, China}

\author{Motoi Kimata}
\affiliation{Advanced Science Research Center, Japan Atomic Energy Agency, Tokai, Ibaraki 319-1195, Japan}



\author{Sofie S{\o}by Leiszner}
\affiliation{Department of Chemistry, Aarhus University, Langelandsgade 140, 8000 Aarhus C, Denmark}

\author{Nhu-Quynh Thi Phan}
\affiliation{Department of Chemistry, Aarhus University, Langelandsgade 140, 8000 Aarhus C, Denmark}

\author{Michael Anthony Quintero}
\affiliation{Department of Chemistry, Aarhus University, Langelandsgade 140, 8000 Aarhus C, Denmark}

\author{Frej S{\o}ren Rattenborg}
\affiliation{Department of Chemistry, Aarhus University, Langelandsgade 140, 8000 Aarhus C, Denmark}

\author{Denis Ar\v{c}on}
\affiliation{Jozef Stefan Institute, Jamova cesta 39, 1000 Ljubljana, Slovenia}
\affiliation{Faculty of Mathematics and Physics, University of Ljubljana, Jadranska cesta 19, 1000 Ljubljana, Slovenia 6}

\author{Bo Brummerstedt Iversen}
\affiliation{Department of Chemistry, Aarhus University, Langelandsgade 140, 8000 Aarhus C, Denmark}

\author{Katsumi Tanigaki}
\affiliation{BAQIS, Bld. 3, No.10 Xibeiwang East Rd., Haidian District, Beijing 100193, China}


\date{\today}

\maketitle

\tableofcontents

\section{Scope of analysis and data sources}
\label{data}
The presented analysis are performed on the newly measured sample au-sa2 and published data of other samples \cite{ogasawara2022,jansa2021,ogasawara2021,huynh2019}.
The analyses using quantum interference and orbital shrinkage, the results of the analyses, and the discussion of $\PT$-symmetry are the original findings of this work.
The temperature and magnetic field dependencies of the transport properties of \bmb{} are consistent across all measured samples.

\section{Synthesis and Sample Quality}
\label{sec:synthesis}
All \bmb{} single crystals were synthesized by flux method using Bi-flux at Advanced Institute for Materials Research (AIMR), Tohoku University, Japan following a method outline in Ref.~\onlinecite{saparov2013}.
A mixture of Ba, Mn, and Bi with the molar ratio of Ba:Mn:Bi = 1:2:5 was loaded into a Al$_2$O$_3$ crucible and put in a quartz ampule inside an Ar-filled glovebox.
Another crucible filled with silica wool was then put in the inverted position on top of the chemical-containing crucible.
The ampule was then sealed under vacuum or partial pressure of Helium, put into an electric box furnace.
The heating program started by ramping the temperature to $\SI{1000}{\degreeCelsius}$, following by dwelling for $\SI{10}{\hour}$, then cooling slowly at the rate of $\SI{2}{\degreeCelsius \per \hour}$ to $\SI{500}{\degreeCelsius}$.
The ampule was then hot-decanted with the help of a centrifugal machine to remove the excess fluxes.
Large crystals of centimeter size and mirror $ab$ surface were obtained.
The crystals were stored in Ar-filled glovebox or evacuated quartz ampules and only extracted for measurements.
All measured crystals exhibit consistency in both quality and physical properties.

The chemical compositions of both as-grown crystals and transport samples were confirmed by energy dispersive X-ray spectroscopy combined with scanning electron microscopy measurements at AIMR and Department of Chemistry, Aarhus University, Denmark (AU).
The measurements indicate that the compositions are close to the stoichiometric ratio $1:2:2$ in all samples.

\begin{figure}
  \centering
  \includegraphics[width = .3\textwidth]{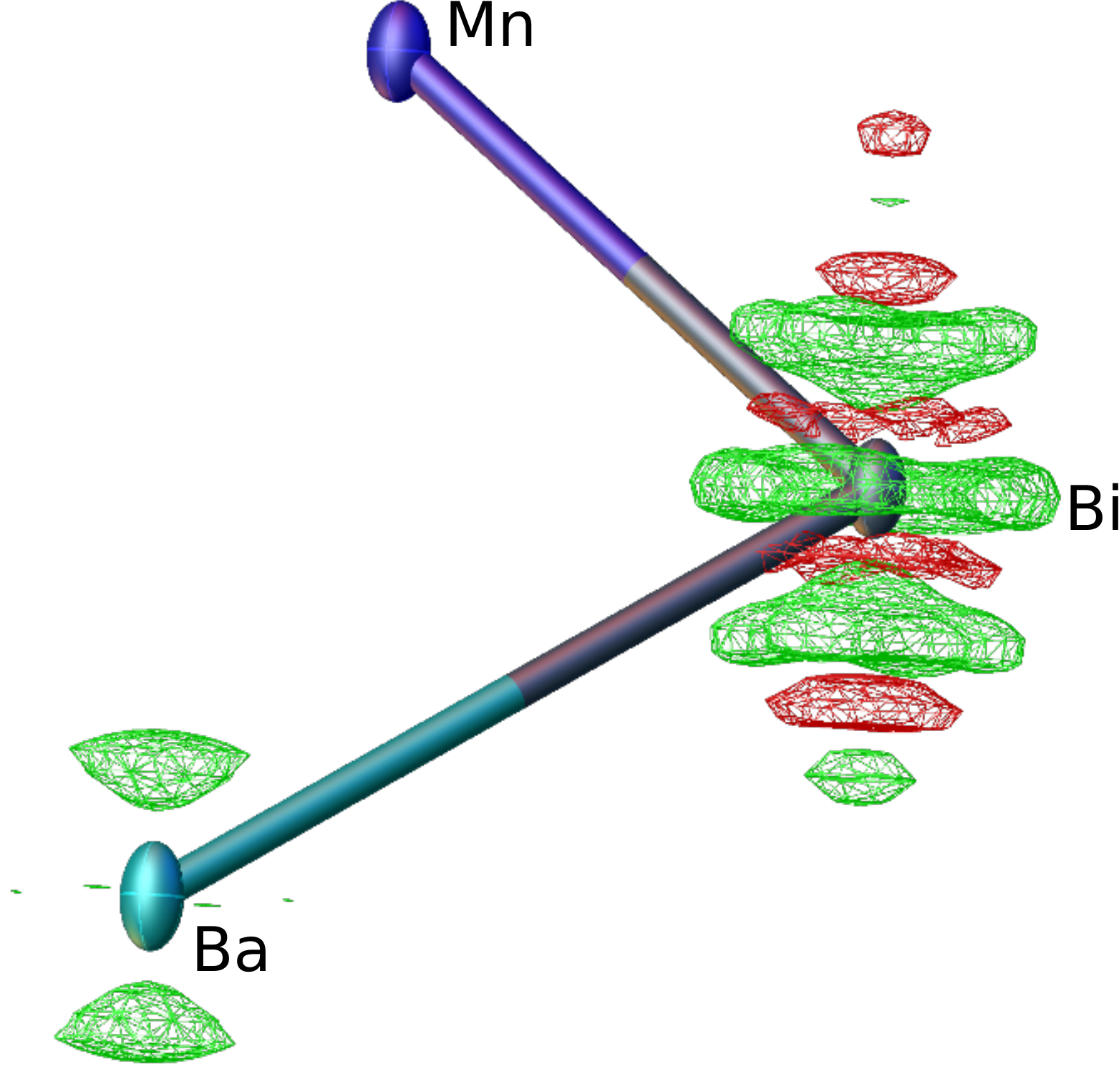}
  \caption{%
    Residual density at contour level $\pm\SI{5}{\electron\per\cubic\angstrom}$.
    Anharmonic thermal motion near bismuth gives rise to the large minimum and maximum residuals.
    The data quality is considered not sufficient for refinement of Gram-Charlier parameters to account for these effects.
  }
  \label{fig:residual}
\end{figure}

Single crystal X-ray diffraction data were collected at $\SI{100}{\kelvin}$ on a Rigaku Synergy-S diffractometer equipped with a HyPix-Arc $\SI{100}{\degree}$ detector installed at AU.
Data reduction, unit cell determination, integration and absorption correction were performed using CrysAlisPro \cite{Crysalis}.
The structure was solved with SHELXT \cite{sheldrick2015} and refined using ShelXL \cite{sheldrick2015a} within the Olex2 interface \cite{dolomanov2009}.
The refined structure confirmed the phase assignment of \bmb{} \cite{calder2014,saparov2013} and was refined to a final $R$-factor of $\SI{7.03}{\percent}$ up to a resolution of $\SI{1.0}{\per\angstrom}$.
Table~\ref{tab:SCXRD} and Fig.~\ref{fig:residual} show the details of refinement and resulted residual density of electron, respectively.
The refined structure was deposited on CCDC database with the deposition number 2588781.

\renewcommand{\arraystretch}{1.2}
\setlength{\tabcolsep}{12pt}

\begin{table}
  \caption{Crystallographic information and data quality parameters for single crystal X-ray diffraction data of \bmb{}.}
  \label{tab:SCXRD}
   \begin{tabular}{l|l}
    \hline
     Chemical formula & \bmb{} \\
     Formula weight $(\unit{\gram\per\mol})$ & $665.16$ \\
     Crystal system & Tetragonal\\
     Space group & $I4/mmm$ \\
     Crystal dimensions $(\unit{\micro\meter})$ & $720\times 510 \times 100$ \\
     Temperature & $\SI{100}{\kelvin}$\\
     Wavelength $(\unit{\angstrom})$ & $0.56087$ \\
     $a\,(\unit{\angstrom})$ & $4.4712(1)$ \\
     $c\,(\unit{\angstrom})$ & $14.6458(4)$ \\
     $V\,(\unit{\cubic\angstrom})$ & $292.793(12)$\\
     $Z$ (number of formula units per cell) & $2$ \\
     $(\sin\theta / \lambda)_{\mathrm{max}}\,(\unit{\angstrom}^{-1})$ & $1.00$ \\
     $R_{\mathrm{int}}\,(\unit{\percent})$ & $30.25$ \\
     Average redundancy & $41.6$ \\
     Completeness $(\unit{\percent})$& $99.8$ \\
     $R(F)\,(\unit{\percent})$ & $7.03$ \\
     Goodness-of-fit & $1.178$ \\
     $\Delta\rho_{\mathrm{min}/\mathrm{max}}\,(\unit{\electron\per\cubic\angstrom})$ & $-5.5, \, 18.0$\\
    \hline
  \end{tabular}
\end{table}

\section{Transport properties}
\label{sec:magnetotransport}

\subsection{Temperature dependence of resistivity}
\label{sec:temp-depend-resist}
\begin{figure}
  \centering
  \includegraphics[width = .3\textwidth]{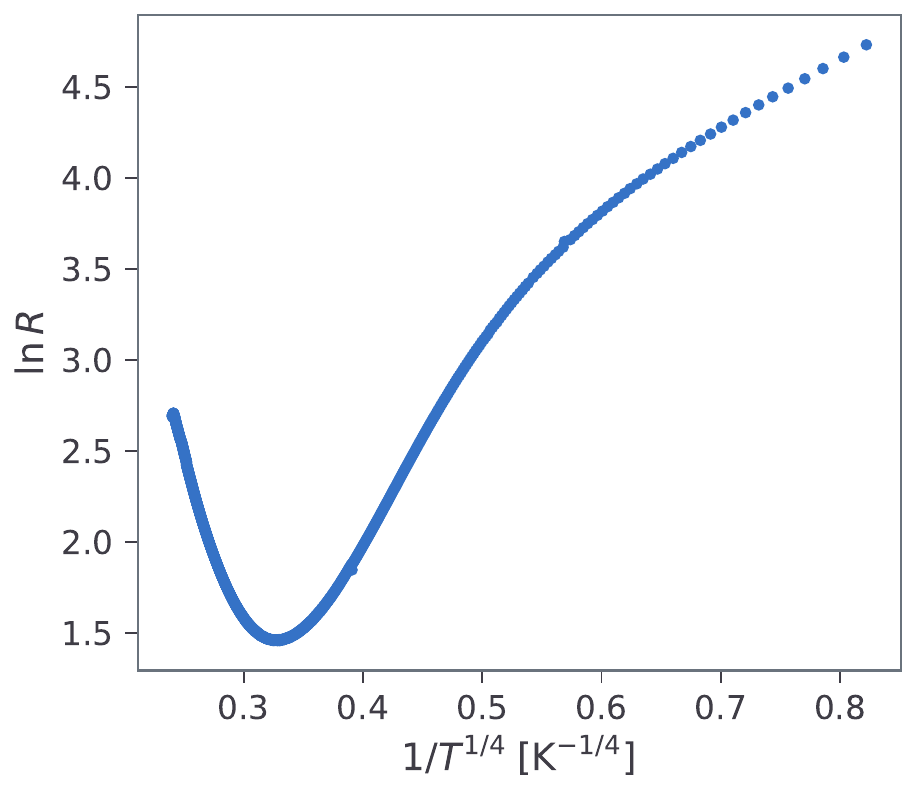}
  \caption{%
    Temperature dependence of the resistivity of \bmb{} (sample au-sa2)
  }
  \label{fig:T-R}
\end{figure}

Fig.~\ref{fig:T-R} the logarithmic resistivity, $\ln R$, of sample au-sa2 is plotted as a function of $T^{-1/4}$.
The sample exhibits a metal-to-insulator transition at $T_{\mathrm{MIT}}\approx \SI{80}{\kelvin}$.
At $T < T_{\mathrm{MIT}}$, the resistivity curve follows an thermal activation rule, $\rho \propto \exp{(-E_{\mathrm{a}}/(k_{\mathrm{B}}T))}$.
When $T\lesssim \SI{25}{\kelvin}$, the transport crosses over to a three-dimensional variable-range-hopping $T$-dependence, $\rho \propto \exp({(T/T_0)^{-1/4}})$.
These behaviors are consistent with previous reports \cite{huynh2019,ogasawara2021,jansa2021,ogasawara2022}.

\subsection{Conductivity and resistivity tensors of \bmb{} in magnetic fields}
\label{sec:cond-resist-tens}

In the main text, we use the notations $s \equiv \sigma(\vec{H}) / \sigma(0)$ and $r \equiv \rho(\vec{H})/ \rho(0)$, and $s$ is calculated directly from the experimentally measured $r$ data using a simple relation between conductivity $\sigma$ and resistivity $\rho$,
\begin{align}
  \label{eq:sigma-rho-def}
  \sigma &\approx \frac{1}{\rho} \,,
\end{align}
and thus $s = 1/r$.
The validity of this relation is justified as follows.

$\bm{\sigma}$ and $\bm{\rho}$ are tensors with non-zero off-diagonal components corresponding to Hall effects in magnetic fields.
However, the approximation given by Eq.~\ref{eq:sigma-rho-def} is good when Hall resistivity is much smaller than the resistivity.
In the case of \bmb{}, the Hall resistivity is about two order of magnitudes smaller than the resistivity  \cite{ogasawara2022,ogasawara2021}.

The resistivity tensor $\bm{\rho}$ of \bmb{} can be written as,
\begin{align}
  \label{eq:rho-tensor}
  \bm{\rho} &= %
              \begin{pmatrix}
                \rho_{aa} & -\rho_{ba} & -\rho_{ca} \\
                \rho_{ba} & \rho_{aa} & -\rho_{ca}  \\
                \rho_{ca} & \rho_{ca} & \rho_{cc}
              \end{pmatrix}\,,
\end{align}
where $\rho_{aa} = \rho_{bb}$ and $\rho_{ca} = \rho_{cb}$ due to tetragonal symmetry, and $\rho_{ij} = -\rho_{ji}$ for $i \neq j$ due to Onsager's relation.
The determinant of $\bm{\rho}$ is,
\begin{align}
  \label{eq:det-rho}
  |\bm{\rho}| &= \rho_{aa}^2\rho_{cc} + 2 \rho_{ca}^2 \rho_{aa} + \rho_{ba}^2\rho_{cc}  \nonumber\\
           {} &\approx  \rho_{aa}^2\rho_{cc}\,.
\end{align}
In \bmb{}, $\rho_{cc} \approx 2\rho_{aa}$ and $|\rho_{ij}| \propto 10^{-2} \rho_{ii}$; therefore the terms with $\rho_{ij}^2$ with $i \neq j$ can be ignored.
Because $\bm{\sigma} = (\bm{\rho})^{-1}$, we have,
\begin{align}
  \label{eq:sigma_aa}
  \sigma_{aa} &= \frac{1}{|\bm{\rho}|}(\rho_{aa}\rho_{cc} + \rho_{ca}^2) \nonumber \\
             &\approx \frac{1}{\rho_{aa}}\,,
\end{align}
and,
\begin{align}
  \label{eq:sigma_aa}
  \sigma_{cc} &= \frac{1}{|\bm{\rho}|}(\rho_{aa}^2 + \rho_{cb}^2) \nonumber \\
             &\approx \frac{1}{\rho_{cc}}\,,
\end{align}
as presented in the main text.

\subsection{Angular dependence of magnetoresistance}
\label{sec:MRvstheta}

\begin{figure}
  \centering
  \includegraphics[width = \textwidth]{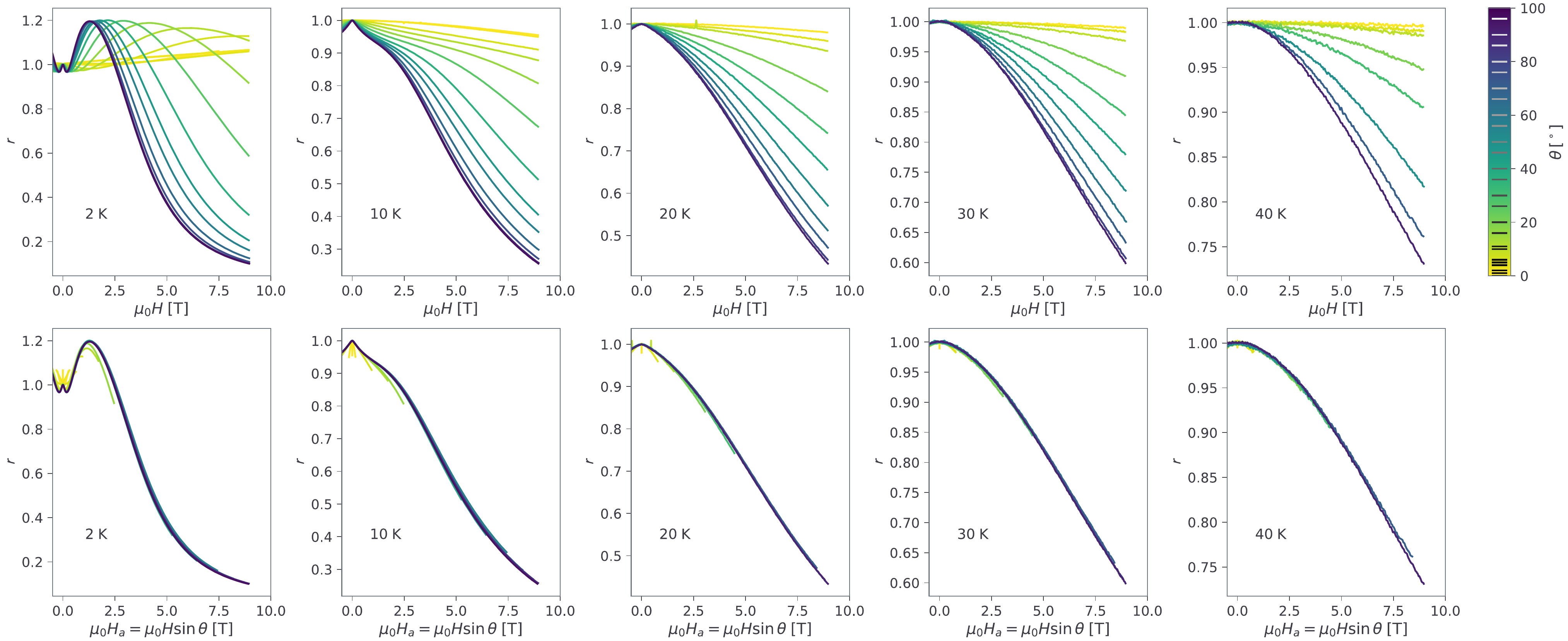}
  \caption{%
    Angular dependencies of the $r$-curves measured at various temperatures.
    The sample is au-sa2.
    The upper row shows $r$-curves measured at different $\theta$-angles as a function of total field strength $H$.
    The lower row shows the scaling of the same data on the transverse field strength $H_a = H\sin\theta$.
    The colors indicate the $\theta$-angles.
  }
  \label{fig:MRvstheta}
\end{figure}

Fig.~\ref{fig:MRvstheta} displays the dependencies on the $\theta$-angle of the $r$-curves measured on sample au-sa2 at various temperatures.
The magnetic field $\vec{H}$ was rotated in the $ac$-plane, whereas the electric current $\vec{j}$ was parallel to the $\hat{a}$-axis.
The first row of the figure shows the $r$-curves collected at various angles.
The second row shows that by plotting the same data as a function of the in-plane field strength $H\sin\theta$, the complex curves collapse into a single form at each temperatures.
The curves start to deviate from the scaling law at $|\theta| < \SI{10}{\degree}$.

\subsection{Temperature dependence of the magnetoresistance}
\label{sec:MRvsT}
The magnetoresistance ($\mathrm{MR}$) of \bmb{} is governed mainly by quantum interference (QI) and radius shrinkage of the localized wavefunction (SWF) mechanisms.
These two mechanisms yield opposite signs of competing $\mathrm{MR}$ components and have different temperature dependencies.
The negative $\mathrm{MR}$ (positive magnetoconductance $\mathrm{MC}$) is caused by QI.
This negative MR is robust and remains observable at $T \lesssim \SI{100}{\kelvin}$.
On the other hand, SWF induces a positive $\mathrm{MR}$ component which fades out quickly with increasing $T$.
The overall $\mathrm{MR}$ therefore shows three regimes in its temperature dependence as follows.
\begin{enumerate}
\renewcommand{\labelenumi}{(\arabic{enumi})}
\item \textbf{Region 1, $T \lesssim \SI{10}{\kelvin}$:}
  the dephasing effects of temperature are weak.
  Both QI and orbital shrinkage are strong, resulting a complex $r$-curve with both negative and positive $\mathrm{MR}$.
  The QI-induced negative $\mathrm{MR}$ starts from the zero-field limit and continues at higher field into a saturation.
  The SWF-induced positive $\mathrm{MR}$ appears as a hump in the field range $\SI{0.1}{\tesla} < \mu_0 H_{ab} < \SI{3}{\tesla}$.
  The magnitude of the SWF-$\mathrm{MR}$ decays quickly with increasing temperatures.
  At $T\approx \SI{10}{\kelvin}$, the positive $\mathrm{MR}$ hump becomes a shoulder in the overall negative $\mathrm{MR}$ curve.

  In this region, the positive $\mathrm{MR}$ component exhibit the dependence $\ln{r} \propto (\mu_0 H_{ab})^2$.
  The negative $\mathrm{MR}$ exhibit the QI-law $\ln s = (\mu_0 H_{ab})^{\alpha}$.
\item \textbf{Region 2, $\SI{10}{\kelvin} \lesssim T \lesssim \SI{25}{\kelvin}$:}
  The SWF shoulder continues to decay.
  At $T\approx \SI{25}{\kelvin}$, this feature completely disappears.
  On the other hand, the QI effects is weaken, and the saturation at high magnetic fields remains clear.
\item \textbf{Region 3, $T \gtrsim \SI{25}{\kelvin}$:}
  The negative $\mathrm{MR}$ is weaken by dephasing effects but remains clear untill $T > 100\,\mathrm{K}$.
\end{enumerate}

The $s$-curves in region 3 are govern by only QI-mechanism and allow a direct estimation of $\alpha$.
In this region, $\ln(\ln s)$ is linear with $\ln(\mu_0H_a)$ with the slope $\alpha$, as shown in Fig. 3(a) of the main text.
Linear fittings in this $T$-region on all samples yield an average $\bar{\alpha} \approx 1.89$.
In $T$-regions 1 and 2, the estimation of $\alpha$ is more difficult due to the presence of the positive $\mathrm{MR}$-component.
Nevertheless, the linear relation between $\ln(\ln s)$ and $\ln(\mu_0 H_a)$ is clear in the low field regions, and the estimated values of $\alpha$ are nearly identical to $1.89$.
Plotting $\ln s$ against $(\mu_0 B)^{1.89}$ also shows a linear relation, which is helpful in validating the QI-mechanism.
Finally, in $T$-region 1, the positive magnetoresistance ($r > 1$) generally obeys $\ln{r} \propto (\mu_0 H_a)^2$.


\section{Symmetry considerations}
\label{sec:symm-cons}
We consider the following Hamiltonian of a $\PT$-symmetric AFM placed in magnetic field $\vec{H}$ from an empirical point of view,
\begin{align}
  \label{eq:H-AFM}
  \mathcal{H}^{\mathrm{AFM}} &= (\mathcal{H}^{\mathrm{AFM}}_0  + \mathcal{H}^{\mathrm{AFM }}_{\mathrm{Z}} ) + \mathcal{H}^{\mathrm{AFM}e} + \mathcal{H}^{e}_{\mathrm{orb}} + \mathcal{H}^{e}_{\mathrm{Z}}\,.
\end{align}
Here $\mathcal{H}^{\mathrm{AFM}}_0$ describes the unperturbed, $\mathcal{PT}$-symmetric AFM state and $\mathcal{H}^{\mathrm{AFM}}_{\mathrm{Z}}$ is the Zeeman interaction between the AFM sublattices and $\vec{H}$.
$\mathcal{H}^{\mathrm{AFM}e}$ models the interactions between the AFM background and the itinerant electrons.
The last two terms stand for the interactions between $\vec{H}$ and the conduction electrons.
$\mathcal{H}^{e}_{\mathrm{orb}}$ is the orbital coupling,
\begin{align}
  \label{eq:Horb}
  \mathcal{H}^{e}_{\mathrm{orb}} &= \frac{1}{2m}(\vec{p} + e\vec{A})^2\,,
\end{align}
where $m$, $\vec{p}$, and $-e$ are the (effective) mass, momentum, and charge of the electron, respectively.
$\mathcal{H}^{e}_{\mathrm{Z}}$ is the Zeeman interaction,
\begin{align}
  \label{eq:HZeeman}
  \mathcal{H}^{e}_{\mathrm{Z}} &= \mu_{\mathrm{B}} \vec{B} \cdot \mathbf{g} \cdot \vec{S}\,,
\end{align}
where $\mathbf{g}$ is a tensor, and
$\mu_{\mathrm{B}}$, $\vec{B}$, and $\vec{S}$ are Bohr magneton, magnetic field, and electron's spin, respectively.

$\Trev$-symmetry, and thus $\PT$-symmetry, are formally broken by any non-zero magnetic field of general direction via its interactions with the AFM sublattices and the itinerant electrons.
Specifically, on the sector of itinerant electrons, $\mathcal{H}^{e}_{\mathrm{orb}}$ alters the momenta and the phase of the electrons while preseves their spin states, leading to Landau quantization and orbital diamagnetism in the equilibrium state.
In charge transport, it causes ordinary Hall effects, and the acquired Aharonov-Bohm phases underlie characteristic magnetoconductances in weak (anti-) localization and Anderson insulating states \cite{aharonov1959,hikami1980,kawabata1980a,nguyen1985,medina1990,medina1992,medina1996}.
The Zeeman interaction $\mathcal{H}^{e}_{\mathrm{Z}}$ splits the spin degeneracy and leads to Pauli paramagnetism.
In paramagnets hosting topologically non-trival band structure, this splitting can drive the system into Weyl semimetallic states \cite{cano2017}.
In the AFM sector, $\mathcal{H}^{\mathrm{AFM}}_{\mathrm{Z}}$ breaks $\PT$-symmetry by reorganizing the AFM magnetocrystalline structure.
This term is anisotropic and its $\PT$-breaking strength can be quantified via the magnetic susceptibility $\bm{\chi}$ when the Pauli and orbital contributions are negligible.
Various theoretical studies have investigated the breaking of $\PT$-symmetry by external magnetic fields at different limits of spin--orbit interaction and the $\mathcal{H}^{\mathrm{AFM}e}$ coupling \cite{ramazashvili2008,yanase2014,watanabe2020}.

How sucseptible the AFM system is to each symmetry breaking term depends strongly on the unperturbed $\mathcal{H}^{\mathrm{AFM}}_{0}$ and the strength of $\mathcal{H}^{\mathrm{AFM}e}$.
In the current case, \bmb{} exhibits a strong easy-axis AFM anisotropy and a Anderson insulating ground state.
The localized conduction electrons come from an Mn $d_{xy}$-derived band, whereas the other $d$-orbitals shows strong Mott-like behavior with the AFM order \cite{ogasawara2022,jansa2021,craco2018}.
This orbital dependence suggests a strong coupling between the conduction electrons and the AFM background.
On the the other hand, the strong localization largely suppresses $\mathcal{H}^{e}_{\mathrm{Z}}$, whereas  $\mathcal{H}^{e}_{\mathrm{orb}}$ remains effective as seen by small but finite Hall effects in $\vec{H}\parallel\hat{c}$ \cite{ogasawara2021,huynh2019}.
However, measurements, which employed various transport geometries, indicate that the effect of this term is small.
For example, in the configuration where $\vec{H} \parallel \hat{c}$ and $\vec{j} \parallel \hat{a}$, with $\vec{j}$ is the electric current, the Lorentz force arising from $\mathcal{H}^e_{\mathrm{orb}}$ is maximized.
In paramagnetic Anderson insulator, this ``transverse'' transport geometry also yield the largest quantum interference magnetoconductance \cite{medina1996}.
By contrast, the same geometry give rise to the smallest the magnetoconductance in \bmb{}, and the magnetoconductance is maximized whenever $\vec{H}$ is parallel to the $ab$-plane, regardless of the orientation of $\vec{j}$ \cite{ogasawara2022,ogasawara2021,huynh2019}.

The observed anistrotropy and $H_{ab}$-dependencies of the magnetoconductane in \bmb{} show that the effects of breaking $\PT$-symmetry by $\vec{H}_{ab}$ are identical to removing an \emph{effective} $\Trev$-symmetry protecting the localization state.
Interestingly, whereas $\PT$-symmetry is broken via Zeeman interaction $\mathcal{H}^{\mathrm{AFM}}_{\mathrm{Z}}$, the induced effects, i.e., the modification of quantum interference and the shrinkage of wavefunction radius, can be represented by an effective orbital coupling $e\vec{A}_{\mathrm{eff}}$.
The formulation of $e\vec{A}_{\mathrm{eff}}$ in terms of the hopping motions of localized electrons across canted AFM sublattices is beyond the scope of this work.

\bibliographystyle{apsrev4-2}
\bibliography{./BMPn_bibtex.bib}

